\documentclass{article}

\PassOptionsToPackage{round}{natbib}
\usepackage[final]{neurips_2024}

\usepackage{amsmath}

\usepackage[utf8]{inputenc} 
\usepackage[T1]{fontenc}    
\usepackage{hyperref}       
\usepackage{cleveref}       
\usepackage{url}            
\usepackage{booktabs}       
\usepackage{amsfonts}       
\usepackage{nicefrac}       
\usepackage{microtype}      
\usepackage{xcolor}         
\usepackage{graphicx} 

\usepackage{algorithm}
\usepackage{algpseudocode}
\usepackage{algorithmicx}
\usepackage{caption}

\newcommand{\intellect}{\textsc{intellect-1}}
\newcommand{\intellectinstruct}{\textsc{intellect-1-instruct}}
\newcommand{\primeframework}{\textsc{prime}}

\title{INTELLECT-1 Technical Report}

\author{%
  Sami Jaghouar \\
  Prime Intellect \\
\And
  Jack Min Ong \\
  Prime Intellect \\
\And
  Manveer Basra \\
  Prime Intellect \\
\And
  Fares Obeid \\
  Prime Intellect \\
\AND
  Jannik Straube \\
  Prime Intellect \\
\And
  Michael Keiblinger \\
  Prime Intellect \\
\And
  Elie Bakouch \\
  Hugging Face \\
\And
  Lucas Atkins \\
  Arcee AI \\
\AND
  Maziyar Panahi \\
  Arcee AI \\
\And
  Charles Goddard \\
  Arcee AI \\
\And
  Max Ryabinin \\
  Together AI \\
\And
  Johannes Hagemann \\
  Prime Intellect \\
  \texttt{johannes@primeintellect.ai} \\
}

\begin{document}

{
\begingroup
\begin{figure*}
    \centering
    \includegraphics[width=0.125\textwidth]{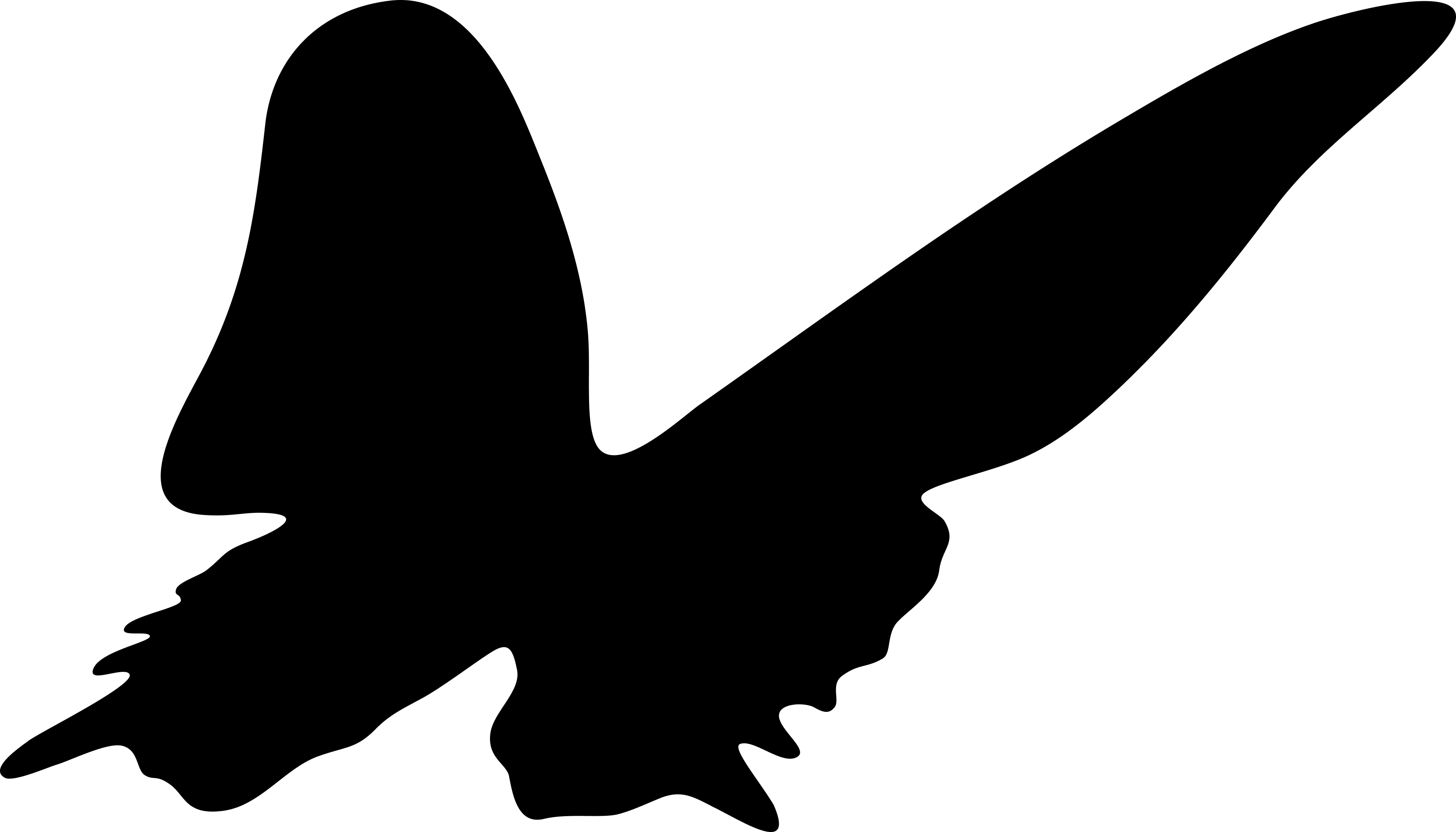}
\end{figure*}
\endgroup
}
\setcounter{figure}{0}

\maketitle

\begin{abstract}
In this report, we introduce \intellect, the first 10 billion parameter language model collaboratively trained across the globe, demonstrating that large-scale model training is no longer confined to large corporations but can be achieved through a distributed, community-driven approach.

\intellect\ was trained on 1 trillion tokens using up to 14 concurrent nodes distributed across 3 continents, with contributions from 30 independent compute providers dynamically joining and leaving the training process, while maintaining 83-96\% compute utilization and 36.2--41.4\% model FLOPS utilization.

We leverage \primeframework, our scalable distributed training framework designed for fault-tolerant, high-performance training on unreliable, globally distributed nodes.
Key innovations in \primeframework\ include the \texttt{ElasticDeviceMesh}, which manages dynamic global process groups for fault-tolerant communication across the internet and local process groups for communication within a node, live checkpoint recovery, kernels, and a hybrid DiLoCo-FSDP2 implementation.

Using \primeframework\ with DiLoCo and our custom int8 all-reduce, we achieve a $400\times$ reduction in communication bandwidth compared to traditional data-parallel training settings while delivering comparable performance.

These results demonstrate the feasibility and promise of training frontier foundation models in a decentralized network of global GPU resources.
\end{abstract}

\newpage

\tableofcontents

\newpage

\section{Introduction}

The rapid scaling of large language models ~\citep{gpt3, palm, gpt4} has demonstrated unprecedented capabilities but also exposed significant challenges in the infrastructure required for their training. Traditional distributed training approaches rely heavily on high-bandwidth interconnects within centralized data centers. However, there is increasing interest in enabling training across geographically distributed nodes, leveraging standard internet connections to pool GPU resources for collaborative model training. This paradigm shift introduces new challenges: network bandwidth between nodes can be up to three orders of magnitude lower than in typical HPC environments, and the pool of training nodes can be dynamic, with nodes joining or leaving the process unpredictably, making system reliability a critical concern~\citep{ryabinin2020crowdsourced}.

In this paper, we present the first large-scale experiment collaboratively training a 10 billion parameter model over 1T tokens across five countries and three continents on up to 112 H100 GPUs simultaneously. We achieve an overall compute utilization of 83\% across continents and 96\% when training exclusively on nodes distributed across the entire United States, introducing minimal overhead compared to centralized training approaches. Notably, nodes run independently for approximately 38 minutes before performing an all-reduce operation, which takes around 1 to 7 minutes depending on the configuration, ensuring efficient utilization while minimizing communication overhead. Our results show that \intellect\ can maintain training convergence and high compute utilization despite severe bandwidth constraints and node volatility, opening new possibilities for decentralized, community-driven training of frontier foundation models.

We introduce the \primeframework\ framework specifically designed for training large models across unreliable, bandwidth-constrained nodes. \primeframework\ introduces several key innovations to address these challenges. At its core is the \texttt{ElasticDeviceMesh}, a novel abstraction that manages both fault-tolerant communication across the internet and efficient local communication within nodes. This hybrid approach combines the benefits of Fully Sharded Data Parallel (FSDP) ~\citep{fsdp} training for intra-node efficiency with the Distributed Low-Communication (DiLoCo) ~\citep{diloco, open_diloco} algorithm for minimal inter-node communication.

To maximize training efficiency in this challenging environment, \primeframework\ implements several crucial optimizations.
We implement an efficient CPU-offloaded version of DiLoCo that does not add more GPU memory overhead than normal distributed training. Our system employs quantization of gradient transfers to 8-bit integers, reducing communication volume by up to $2000\times$ to standard data parallel training by combining int8 quantization of gradients with an outer optimizer synchronization every 500 steps.
Furthermore, \primeframework\ features robust fault tolerance mechanisms, including dynamic node addition and removal, and efficient checkpoint recovery protocols.

Using \primeframework, we successfully collaborated to train \intellect\ on a high-quality dataset comprising 1 trillion tokens. This achievement was made possible through the contributions of 30 independent compute sponsors, including industrial partners and individual supporters, who provided H100 nodes from around the globe. The training process was accompanied by a public dashboard\footnote{Public training dashboard: \url{https://app.primeintellect.ai/intelligence}}, allowing real-time monitoring of performance.

We open-source the \intellect\ base model, intermediate checkpoints, pre-training data, post-trained checkpoints, post-train data at \href{https://huggingface.co/PrimeIntellect/INTELLECT-1}{\texttt{huggingface.co/PrimeIntellect/INTELLECT-1}} and  the \primeframework\ framework at \href{https://github.com/PrimeIntellect-ai/prime}{\texttt{github.com/PrimeIntellect-ai/prime}}.

The remainder of this report is organized as follows: \Cref{prime-framework} provides a detailed overview of the key features of the \primeframework\ framework. \Cref{intellect-1-training} describes the experimental setup for \intellect, compute efficiency analysis, training behavior, post-training techniques, and model evaluation. \Cref{discussion-decentralized-ai} discusses possible implications of decentralized training for the open-source AI ecosystem. Finally, \Cref{conclusion} concludes the report and outlines directions for future work.

\section{Prime Framework: Enabling Scalable Decentralized Training}
\label{prime-framework}

Modern distributed training of large language models typically relies on high-bandwidth interconnects, with InfiniBand networks providing up to 3.2 Tb/s communication speeds between nodes.
However, when training needs to be distributed across geographically distant locations, such high-bandwidth connections are unavailable.
In our setting, the maximum available bandwidth between nodes is limited to 500 Mb to 4 Gb/s --- three orders of magnitude lower than typical HPC environments.
This severe bandwidth constraint makes communication the primary bottleneck for distributed training, necessitating new approaches to efficient model synchronization.

To address this constraint, we propose an implementation of the Distributed Low-Communication (DiLoCo) algorithm~\citep{diloco}. DiLoCo has been shown to enable training across poorly connected devices while maintaining convergence comparable to traditional distributed training~\citep{open_diloco}. For further bandwidth reduction, we experiment with performing 8-bit quantization of pseudo-gradients, which combined with DiLoCo's reduced synchronization frequency can provide up to $2000\times$ reduction in communication volume compared to standard data parallel training while maintaining model quality.

DiLoCo performs a version of Local SGD~\citep{stich2018local}, where each independent worker executes $H$ local optimization steps using AdamW before synchronizing through an outer optimization step using Nesterov momentum. The key insight is that communication between workers is required only every $H$ steps (typically up to 500), dramatically reducing bandwidth requirements compared to traditional data parallel training that communicates gradients at every step.

\begin{algorithm}
\caption{DiLoCo Algorithm} \label{alg:diloco}
\begin{algorithmic}[1]
\Require Initial model $\theta^{(0)}$
\Require $k$ workers
\Require Data shards $\{\mathcal{D}_1, \dots, \mathcal{D}_k\}$
\Require Optimizers \texttt{InnerOpt} and \texttt{OuterOpt}
\For{outer step $t = 1 \dots T$}
    \For{worker $i = 1 \dots k$}
        \State $\theta_i^{(t)} \gets \theta^{(t-1)}$
        \For{inner step $h = 1 \dots H$}
            \State $x \sim \mathcal{D}_i$
            \State $\mathcal{L} \gets f(x, \theta_i^{(t)})$
            \State \Comment{Inner optimization}
            \State $\theta_i^{(t)} \gets \texttt{InnerOpt}(\theta_i^{(t)}, \nabla\mathcal{L})$
        \EndFor
    \EndFor
    
    \State $\Delta^{(t)} \gets \frac{1}{k} \sum_{i=1}^k (\theta^{(t-1)} - \theta_i^{(t)})$ \label{lst:pseudo_grad} \Comment{Averaging outer gradients}
    \State $\theta^{(t)} \gets \texttt{OuterOpt}(\theta^{(t-1)}, \Delta^{(t)})$ \Comment{Outer optimization}
\EndFor
\end{algorithmic}
\end{algorithm}

In this section, we detail our implementation that enables efficient DiLoCo training across bandwidth-constrained environments.

\subsection{Efficient DiLoCo Implementation}

\primeframework\ offers an efficient implementation of DiLoCo that maintains the same GPU memory footprint as standard distributed training. Pseudo-gradient computation is offloaded to the CPU, followed by a custom IP-based ring all-reduce~\citep{allreduce} implementation, which is required for efficient aggregation of pseudo-gradients. While the time taken for the outer all-reduce operation can be up to multiple minutes depending on the specific connectivity of the peers, its infrequent nature still allows for high overall GPU utilization.

The host-offloaded outer optimization process consists of the following steps:

\begin{enumerate}
    \item Compute pseudo-gradients and perform all-reduce synchronization
    \item Execute the outer optimizer step given the reduced pseudo-gradients on CPU
    \item Transfer the updated weights back to GPU workers
    \item Retain the model weights on CPU for pseudo-gradient delta computation of the next step
\end{enumerate}

This implementation requires additional CPU memory to store the model parameters, outer optimizer states, and pseudo-gradients. While this represents an increased memory burden, modern GPU servers typically have abundant system memory, making this trade-off acceptable.

\subsection{Int8 Ring-All-Reduce}

To lower the required communication bandwidth of the pseudo-gradient commmunication, we implement a specialized int8 quantized ring-all-reduce \citep{allreduce} kernel with fp32 accumulation.

Communicating quantized int8 values instead of the original fp32 values results in a 4 times reduction in communication payload.
We employ the uniform quantization strategy with clipping from ~\cite{hivemind}:

\begin{enumerate}
    \item For each tensor, we compute the mean ($\mu$) and standard deviation ($\sigma$)
    \item The quantization range is then set to $[\mu - 6\sigma, \mu + 6\sigma]$.
    \item This range is uniformly divided into 256 buckets.
    \item The codebook values are determined by computing the average value within each bucket.
\end{enumerate}

This approach effectively handles outliers while preserving the distribution of values within the most significant range of the tensor.

Naive application of the reduce function in the quantized format will result in numeric precision losses.
Specifically, $(Q(a) + Q(b)) \neq Q(a + b)$, where $a$ and $b$ are fp32 values and $Q$ represents the above quantization function.
Therefore, we merely quantize the reduce-terms during transmission while performing the reduction operation in full precision (fp32).

This approach requires significant compute overhead for quantization and dequantization.
We alleviate this by performing pipelined execution of the ring all-reduce, allowing the quantization and dequantization compute overhead to be overlapped with the communication overhead by scheduling them asynchronously.
This hides much of the quantization overhead with the communication time, resulting in a minimal increase in the wall clock time.

In order for our implementation to be fast enough to fully utilize our target maximum bandwidth of 4 Gb/s, we implement custom multithreaded uint8 operations in C++ to perform the quantization and dequantization operations, improving the quantization speed by more than 60 times compared to the PyTorch implementation.
These optimizations prevent quantization from becoming a bottleneck while maintaining the benefits of reduced communication volume from int8 transmissions.

Quantization during training is known to degrade model performance at scale, particularly over extended training periods, due to the emergence and accumulation of outliers ~\citep{llmint8, scaling_law_quant}.
However, our approach differs fundamentally from traditional weight quantization: instead of quantizing the model weights directly, we quantize the pseudo-gradients, which represent the difference between the model states at different timestamps.
We argue that quantizing these temporal differences is inherently more robust than quantizing the weights themselves, as it captures the relative changes in outliers rather than their absolute values.
Furthermore, in the DiLoCo framework, communication reduction through quantization can be balanced against increasing the number of local steps, offering flexibility in optimizing the communication-convergence tradeoff.

\subsection{Hybrid FSDP and DiLoCo}

We implement a hybrid approach combining FSDP (Fully Sharded Data Parallellism) within nodes and DiLoCo across nodes. Each node performs local gather and all-reduce operations via fast intra-node peer-to-peer communication to facilitate FSDP and local data parallelism, while DiLoCo handles the inter-node synchronization over IP. Model parameters and optimizer states remain sharded, with each rank responsible for managing associated system memory state necessary for pseudo-gradient computation and the outer optimizer.

Communication for the all-reduce operation of the outer optimizer is orchestrated such that only
ranks responsible for the same shard communicate. Ranks responsible for different shards of data do not communicate via DiLoCo and will receive the updated weights for other regions only via the gather operations associated with FSDP.

This hybrid approach allows us to leverage both the memory efficiency of FSDP for local training and the communication efficiency of DiLoCo for cross-node synchronization. Within each node, FSDP can take advantage of high-bandwidth GPU interconnects (NVLink/SXM) for efficient intra-node communication, while DiLoCo minimizes the impact of slower inter-node network speeds.

\subsection{Fault Tolerance and Dynamic Node Management}

\begin{figure*}
  \centering
  \includegraphics[width=\linewidth]{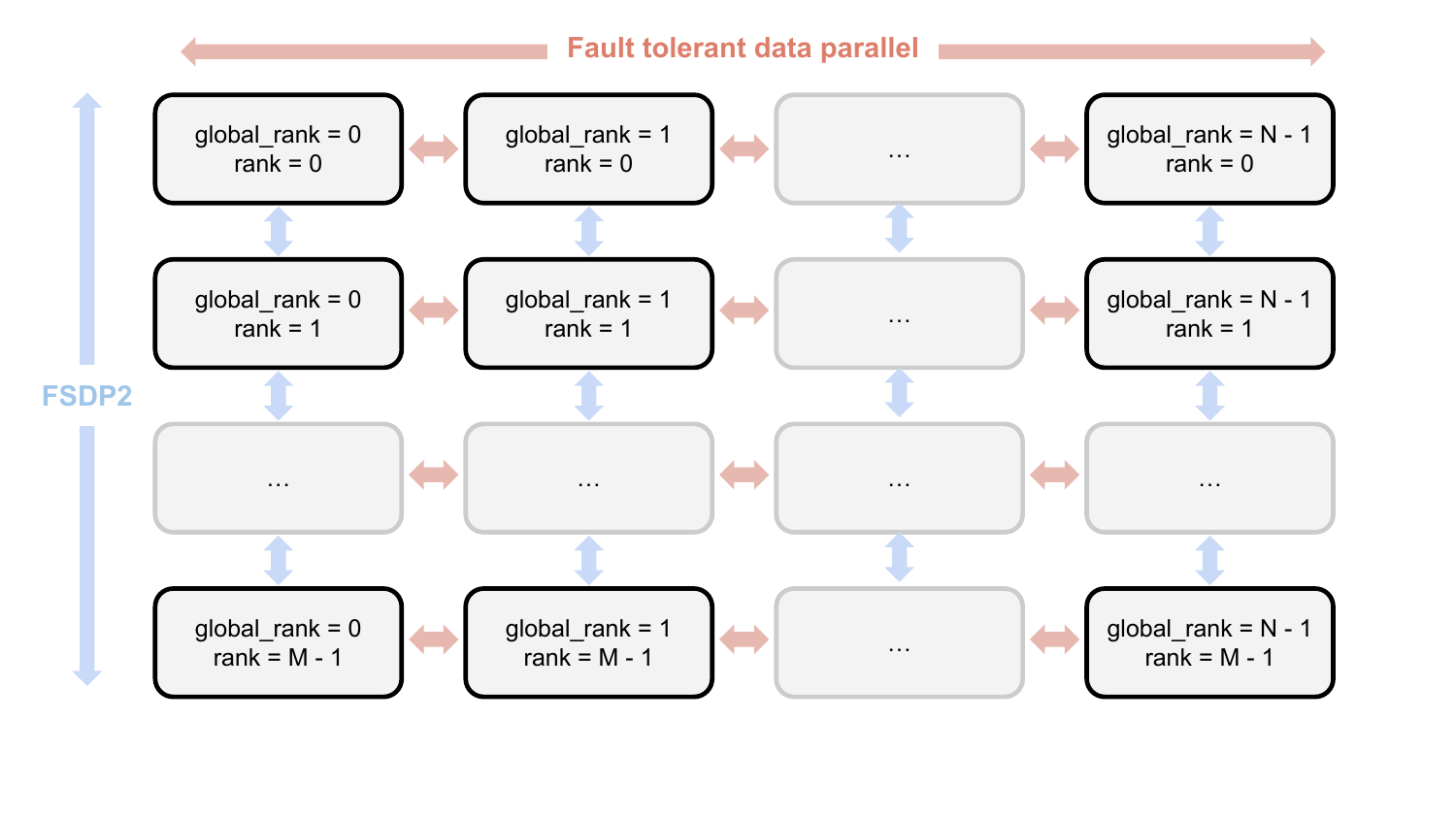}
  \caption{The topology of the \texttt{ElasticDeviceMesh}. Each process in the \texttt{ElasticDeviceMesh} is assigned a local and global rank. The local rank is used by the FSDP process groups, while the global rank is used by an independent fault-tolerant data-parallel process group.}
  \label{fig:topology}
\end{figure*}

This section describes \primeframework's architecture for handling dynamic node participation in distributed training.
We focus on two key capabilities: (1) allowing new nodes to join an ongoing training session without disrupting active nodes, and (2) maintaining training continuity when nodes fail or leave the training run.

\subsubsection{Dynamic Process Groups}

Our implementation focuses specifically on data-parallel training configurations, where the primary collective communication operation is all-reduce.

Data-parallel training is uniquely suited for dynamic world sizes because:
\begin{enumerate}
    \item The all-reduce operation naturally accommodates varying numbers of participants.
    \item Node ranks can be assigned arbitrarily, since the all-reduce operation treats all participants equally.
    \item Model state is replicated across all nodes without complex sharding schemes.
\end{enumerate}

These properties make data-parallel training significantly more flexible than other parallelism strategies when handling dynamic node membership.

\subsubsection{Peer to Peer Checkpoint Transmission}

When a new node joins an ongoing training session, it must synchronize its model and optimizer states with the existing cluster through peer-to-peer checkpoint transfer.
LocalSGD variations offer a significant advantage over data-parallel training approaches, which do synchronization at every step, by being able to overlap this synchronization with the computation of local steps.
Rather than relying on a centralized storage system, we implement two peer-to-peer synchronization strategies:

\begin{enumerate}
    \item Non-blocking synchronization: The joining node directly downloads the checkpoint from any available active peer while training continues. Upon completion, the new node skips the current set of inner steps and joins at the next outer step with zero pseudo-gradients. This approach maximizes cluster utilization by avoiding training interruptions.

    \item Blocking synchronization: Active nodes pause training while the new peer downloads the checkpoint directly from one of the active nodes. This ensures perfect synchronization across all nodes at the cost of temporary training pause.
\end{enumerate}

While the non-blocking approach theoretically offers better cluster utilization, our experiments showed that the blocking strategy was more practical for our setup. Although non-blocking synchronization worked without causing training instability, we observed small loss spikes in the new joiners' initial steps. Given that we added new nodes relatively infrequently (every few days) and peer-to-peer checkpoint transfers completed within 30-60 minutes, we opted for the more conservative blocking approach to ensure maximum training stability.

\subsubsection{Node Removal and Failures}
Nodes may leave the training process for two primary reasons.
A node may leave gracefully through a planned exit, or it might crash unexpectedly.
To handle both scenarios, we implement a heartbeat mechanism, where each running node maintains a subprocess that sends a heartbeat signal to the master key-value store at 2-second intervals. Nodes that fail to send a heartbeat for 6 seconds are automatically evicted from the process group.
In the case of a graceful exit, the node sends a "deathrattle" signal to the master key-value store, triggering immediate removal from the process group without waiting for the timeout.
For crash scenarios, we implement a best-effort attempt to send a deathrattle before the crash occurs.
If this fails, the heartbeat timeout mechanism serves as a fallback, eventually removing the unresponsive node.

\subsubsection{Parallel TCP Stores}
We implemented multiple parallel TCP stores to manage distributed communication, with each store corresponding to a specific replica group.
A replica group consists of nodes sharing the same data-parallel rank, thus containing identical shards of the model, gradient, and optimizer state.
This partitioning approach was designed to reduce network congestion by isolating communication within relevant groups rather than broadcasting all updates across a single shared store.
However, this architecture introduced synchronization challenges.
With separate TCP stores operating independently, maintaining consistent state across all replica groups became more complex.
The isolation that provided performance benefits also made it more difficult to ensure all groups remained properly synchronized, particularly during node failures or dynamic group resizing operations.

\subsubsection{Retries, Timeouts and Edge Cases}
The distributed nature of the system necessitates robust handling of failures during collective operations. When a node fails during an all-reduce operation, we implement a retry mechanism that excludes the failed node and attempts to complete the operation with the remaining healthy nodes.
A particularly challenging edge case emerges when node failure affect only a subset of replica groups. In this scenario, processes running on the same physical node may become desynchronized, leading to misaligned timeout windows. This desynchronization can result in undefined behavior as different processes attempt to proceed with different world views of the available nodes.
The timing issues are exacerbated by the parallel TCP store architecture, as each store may detect and react to failures independently. This can lead to a cascade of timing misalignments across the system. Current timeout values are set empirically based on observed network latencies and failure detection windows, but maintaining consistent timeout behavior across all components remains an open challenge.

\subsection{Networking}

All nodes involved in the training run were connected through a shared Virtual Private Network (VPN) established using Tailscale\footnote{\url{https://tailscale.com/}}, a VPN service based on WireGuard.

We utilize a VPN because our inter-node collective communication library, Gloo\footnote{ \url{https://github.com/facebookincubator/gloo}}, requires all participating processes to be reachable via a private network interface.
Additionally, the VPN ensures secure, encrypted communication between nodes.

Because our code utilizes independent process groups for each rank, we are able to open multiple connections between nodes. This seems to have provided a noticeable increase in the communication bandwidth in our setup.

One issue we observed in the first few days of the training was that the bandwidth between nodes can be quite unreliable.
This means that a high-bandwidth connection might suddenly become slow randomly and for no apparent reason.
This would result in the high-bandwidth sequence we pick for the ring-all-reduce to become low-bandwidth.
We may also experience this issue when nodes leave the training and change the ring topology.

To optimize the performance of global all-reduce operations, we run a background process that continuously measures the bandwidth between nodes.
Using these bandwidth measurements, the process finds the optimal ring-order of nodes such that the minimum bandwidth along the cycle is maximized.
This is achieved by solving a variation of the Traveling Salesperson Problem, where the goal is to find a Hamiltonian cycle (a cycle that visits every node exactly once and returns to the starting node) that maximizes the minimum edge bandwidth in the cycle.
This can be expressed as:
\[
\max_{C \in \mathcal{C}} \min_{(u, v) \in C} w(u, v),
\]

where \( \mathcal{C} \) is the set of all Hamiltonian cycles in the graph \( G = (V, E) \) and \( w(u, v) \) is the bandwidth of the edge \((u, v) \).
The nodes then use this optimized ring-order for all-reduce operations.

During our testing, we observe varying bandwidths when using Tailscale.
Interestingly, there were cases where bandwidth with Tailscale was higher compared to standard internet routing.
However, we also encountered instances where bandwidth was noticeably lower when using Tailscale.
We hypothesize that these variations are influenced by internet routing dynamics.
Tailscale leverages its internal network infrastructure for some connections, which can either outperform or underperform standard internet routing governed by Border Gateway Protocol (BGP).
Factors such as latency, route optimization, and network congestion likely contribute to these observed differences.

Ultimately, our analysis revealed that Tailscale became a bottleneck compared to normal IP-based communication, particularly for intercontinental data transfer. We believe that implementing or leveraging an All-Reduce library that operates over public IP addresses, similarly to Hivemind~\citep{hivemind}, would enhance our training performance.

\subsection{Future work}
The \intellect\ training experiment has demonstrated the need for completely replacing the established collective communications software stack for the realm of internet communication, as intra-datacenter communication is tailored to very different needs that do not translate to global internet communication. Thus, we have started work on our own collective communications library named PCCL (``Prime Collective Communications Library''), which implements a custom TCP-based network protocol called ``Collective Communications over IP'', which will automatically handle peers joining and leaving gracefully, synchronization of designated shared state and bandwidth-aware topology optimization.

\newpage

\section{INTELLECT-1 Training}
\label{intellect-1-training}
\subsection{Experimental Setup}

\begin{figure}[b]
  \centering
  \includegraphics[width=\linewidth]{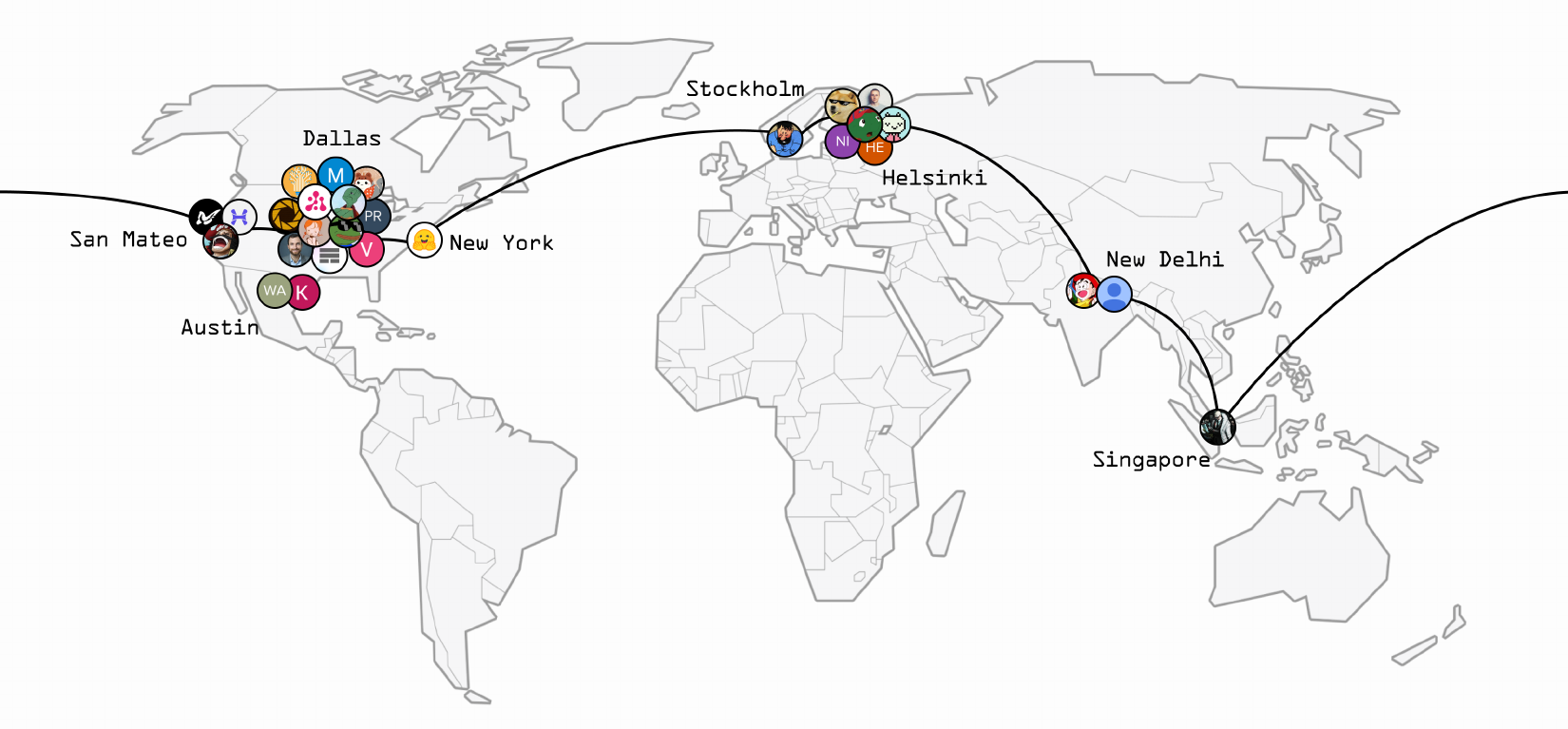}
  \caption{Locations of the nodes by all 30 compute contributors for \intellect. The lines between nodes illustrate the Ring-All-Reduce topology, spanning the whole globe from the US to Europe, Asia, and back to the US.}
  \label{fig:intellect-1-map}
\end{figure}

Using \primeframework, we conducted a public training run of a 10 billion parameter model distributed across up to eight non-colocated data centers spanning three continents (North America, Europe, and Asia). The model is based on the Llama 3 architecture~\citep{llama3}. It was pre-trained with a sequence length of $8,192$ on a total of $1T$ tokens with a high quality data mix (see \Cref{tab:pretraining-dataset} for details).
We employ a weight-space decay (WSD) learning rate scheduler~\citep{hägele2024scalinglawscomputeoptimaltraining} with an inner learning rate of $7.5e^{-5}$ for the AdamW optimizer.
The WSD learning rate scheduler maintains a constant learning rate after an initial warm-up phase and offered us flexibility in the number of tokens we can train on, depending on the number of compute contributions.
During the final 20\% of training, we applied a learning rate cool-down phase along with additional post-training optimizations.
For optimal performance, we chose a conservative number of 100 inner steps between each DiLoCo synchronization, which, on clusters of 8xH100 nodes, took roughly 38 minutes to complete. We quantize the pseudo-gradients to int8, reducing communication requirements by a factor of $400\times$. The average all-reduce synchronization for the DiLoCo outer optimizer using \primeframework\ took approximately 2 minutes for nodes across the United States and 10 minutes for global training, limiting communication between nodes to just 2-10\% of the total training time. The model was trained using an auxiliary max-z-loss \citep{baichuan2} for improved training stability.
For more details about the model configuration, see \autoref{appendix:model-configuration}.

This experimental setup represents the first successful training of a model of this scale across such geographically distributed compute resources with limited network bandwidth.

In the following subsections, we analyze the compute efficiency across different geographical distributions, examine the system's resilience to node changes, detail the pre-training process and results, describe our post-training optimization techniques, and present comprehensive evaluation results against comparable models.

\begin{table}[ht]
\centering
\captionsetup{skip=6pt}
\begin{tabular}{lcc}
\toprule
\textbf{Dataset Name}     & \textbf{Weight (\%)} & \textbf{Weight For Annealing Phase (\%)} \\ \midrule
FineWeb-Edu~\citep{fineweb}              & 55 & 80                  \\ 
FineWeb~\citep{fineweb}                  & 10 & 10                 \\ 
StackV1-popular~\citep{thestack}         & 20 & 10                 \\ 
DCLM-baseline~\citep{dclm}               & 10 &  0                \\ 
OpenWebMath~\citep{openwebmath}          & 5  &  0                \\ \bottomrule
\end{tabular}
\caption{\textbf{Pre-training Dataset Composition:} The model was trained on a diverse mixture of high-quality datasets weighted as follows. As some of these datasets contain shards that contain correlated data, we pre-shuffled the datasets by random sampling from 12 streaming dataset iterators and resharding the dataset. The total number of tokens in our data mix, processed with the Llama-3 tokenizer, consists of over 6 trillion tokens. We adjusted the data mix to prioritize dataset segments with substantially higher loss for the learning rate annealing phase of the training.}
\label{tab:pretraining-dataset}
\end{table}

\subsection{Compute Efficiency Analysis}

We analyze the compute efficiency across three geographical distributions of training nodes: within the United States (USA), between USA and Europe (Transatlantic), and across USA, Europe, and Asia (Global). \Cref{tab:perf_metrics} presents the Model FLOPS Utilization (MFU)~\citep{palm} and timing breakdown for each scenario.

\begin{table}[h]
\centering
\small
\captionsetup{skip=6pt}
\setlength{\tabcolsep}{3pt}
\begin{tabular}{lcccc}
\toprule
\textbf{Scenario} & \textbf{MFU (\%)} & \textbf{Inner step time, min} & \textbf{Median All-Reduce time, s}  & \textbf{Compute Util (\%)} \\
\midrule
Baseline (no comm) & 43.3 & 38 & - & 100 \\
USA & 41.4 & 38 & 103 & 95.7 \\
USA + Europe & 37.1 & 38 & 382 & 85.6\\
Global & 36.0 & 38 & 469 & 83.0 \\
\bottomrule
\end{tabular}
\caption{Performance metrics for training across different geographical configurations. Compute utilization refers to the proportion of time the training is not communicating with other nodes.}
\label{tab:perf_metrics}
\end{table}

The baseline training configuration without communication achieves 43\% MFU, which decreases to 41.4\% for USA-only distribution.
When extending to transatlantic training, MFU drops to 37.1\%, and further decreases to 36.2\% in a global setting.
The computation time for 100 inner steps remains constant at 38 minutes across all scenarios. However, the all-reduce communication time varies significantly: the median is 103 seconds within the USA, increasing to 382 seconds for transatlantic communication, and reaching 469 seconds in the global setting. For reference, checkpoint saving to disk takes 60 seconds. The overhead from CPU-based pseudo-gradient computation and outer optimizer steps is negligible at 5-10 seconds

\begin{figure}[ht]
    \centering
    \includegraphics[width=0.36\linewidth]{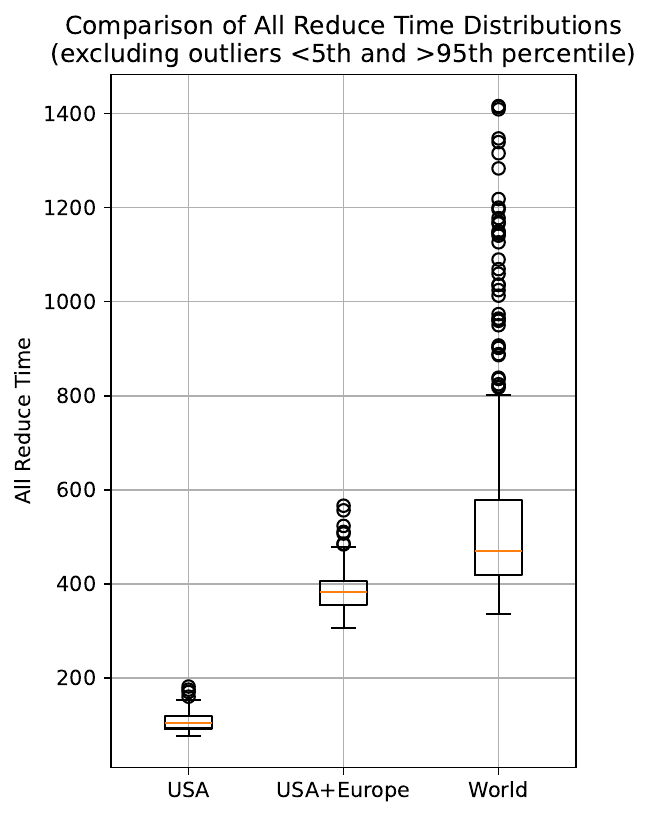}
    \caption{Distribution of all-reduce operation times across different geographical configurations. The variance increases significantly as we move from USA-only to global distribution, indicating less reliable network conditions.}
    \label{fig:box_plot_all_reduce}
\end{figure}

\begin{figure}[ht]
    \centering
    \includegraphics[width=\linewidth]{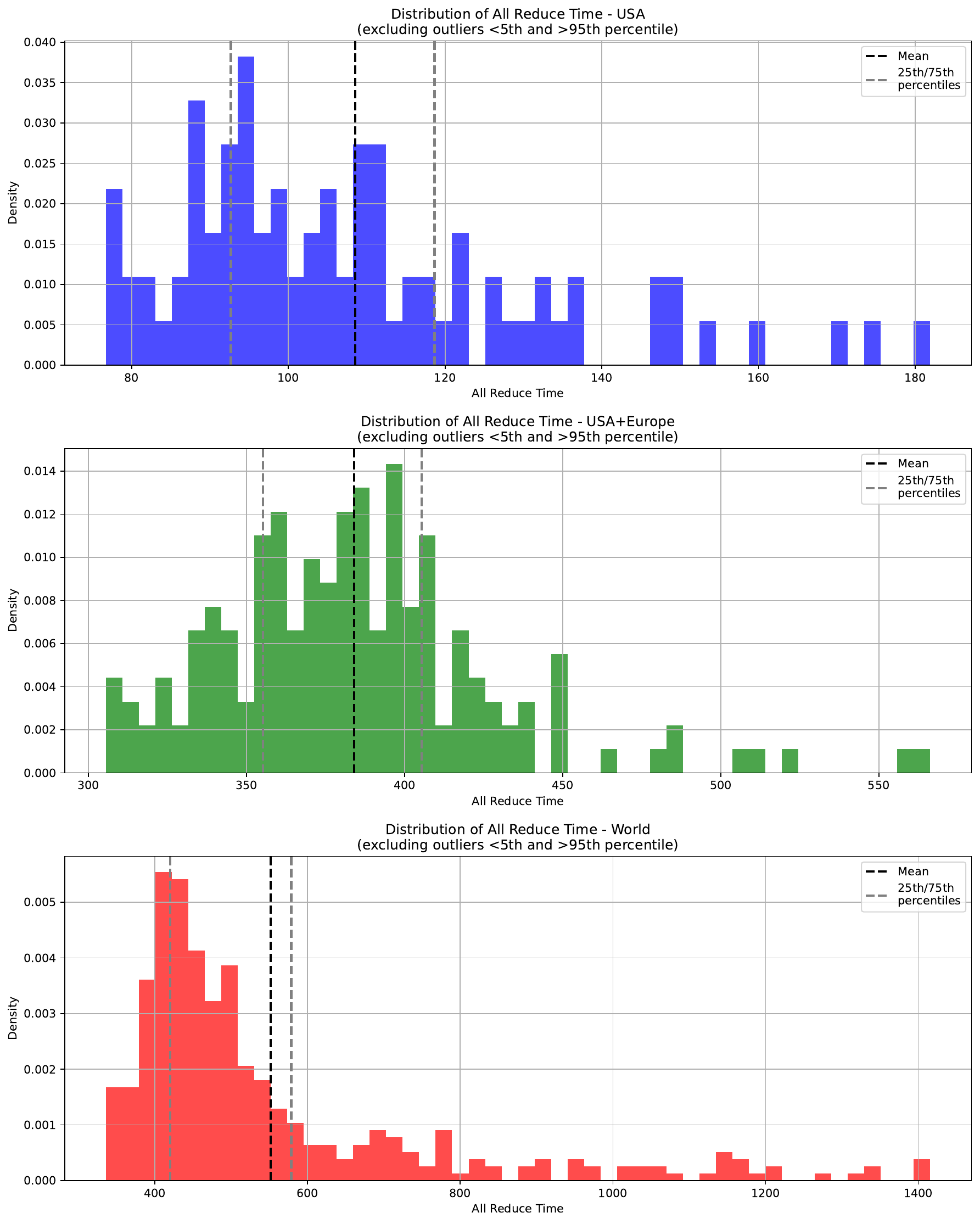}
    \caption{Distribution of all-reduce completion times across different geographical setups. The increasing spread and right-skewed nature of the distributions highlight growing network instability as geographical distances increase. Red represents global, green represents USA and Europe, and blue represents USA-only training.}
    \label{fig:dist_all_reduce}
\end{figure}

Beyond the absolute performance implications of different data types and buffering strategies, our experiments revealed significant variability in network reliability across geographical distributions. As shown in Figure \ref{fig:box_plot_all_reduce}, the variance in all-reduce completion times increases dramatically as we expand from USA-only to global distribution. While USA-only operations show relatively tight bounds, global operations exhibit extensive outliers, with some operations taking up to 2x longer than the median time. Figure \ref{fig:dist_all_reduce} further illustrates this pattern through the probability distributions of completion times, showing increasingly heavy right tails as geographical distance grows. This network instability underscores the importance of robust buffering and quantization strategies to minimize the impact of network variability on training performance.

\subsection{Analysis of Resilience to Node Changes}

\begin{figure}[t]
    \centering
    \includegraphics[width=0.8\linewidth]{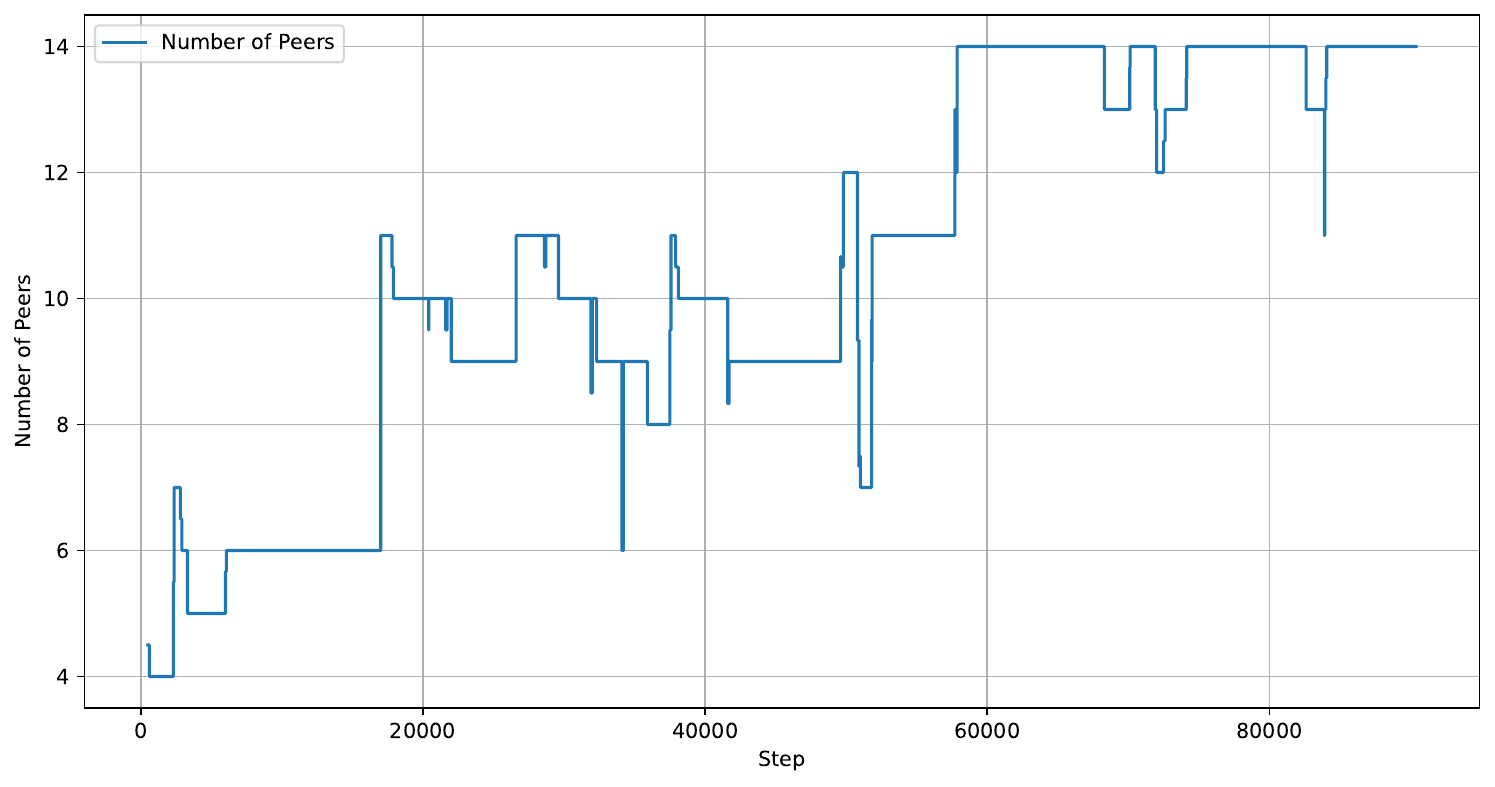}
    \caption{Number of active training nodes over training steps. The graph demonstrates \primeframework's ability to handle dynamic node participation, starting with 4 nodes and scaling up to 14 nodes, while maintaining training stability despite frequent node fluctuations.}
    \label{fig:peer_evolution}
\end{figure}

Our fault-tolerant implementation of DiLoCo demonstrates robust fault tolerance capabilities in distributed training environments. As shown in Figure \ref{fig:peer_evolution}, our training process began with just 4 nodes and gradually scaled up to 14 nodes over time. The system successfully maintained training stability while handling both the gradual onboarding of new nodes and occasional node departure and failures, with the number of active peers fluctuating throughout the training process.

While DiLoCo demonstrates robust adaptation to gradual node changes, we observed that simultaneous loss of multiple nodes (in our case, 4 out of 12) can impact training stability. In this specific case, we opted to resume from a checkpoint. Although such large-scale simultaneous node failures are rare in practice, future work could explore additional mechanisms to handle extreme scenarios where a significant portion of the training fleet is lost at once. 

\subsection{Pre-training}

\begin{figure}[ht]
    \centering
    \includegraphics[width=1\linewidth]{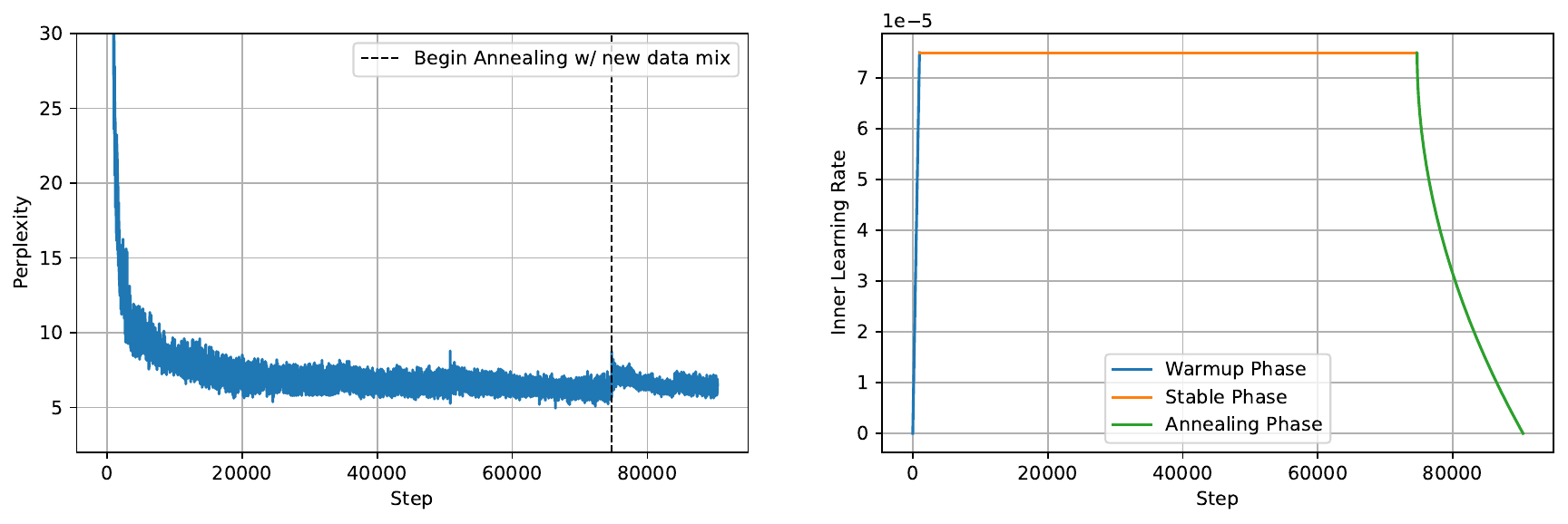}
    \caption{Training dynamics showing model perplexity and learning rate over training steps, including warmup, stable, and annealing phases. The left graph demonstrates the decrease in perplexity from initial training through convergence, while the right graph illustrates the learning rate schedule with distinct warmup, stable, and annealing phases.}
    \label{fig:loss_and_lr}
\end{figure}

The pre-training of \intellect\ for 1 trillion tokens took place over 42 days, from October 10, 2024, to November 22, 2024.
\Cref{fig:loss_and_lr} presents the loss curve over the training steps, demonstrating stable convergence despite significant dynamic on-/off-boarding of compute resources. Around 80\% of the way through training (step $74{\small,}700$), the learning rate annealing phase began. At this stage, we adjusted the data mix to prioritize dataset segments with substantially higher loss, thereby reducing the weight of the Stack~\citep{thestack} and OpenWebMath~\citep{openwebmath} in the mix. This adjustment accounts for the increased loss observed at that point, as the loss on code was significantly lower than text. 

\subsection{Post-training}

After completing the globally distributed pretraining phase, we applied several post-training techniques to enhance \intellect's capabilities and task-specific performance. Our post-training methodology consisted of three main phases.

First, we conducted an extensive series of 16 Supervised Fine-Tuning (SFT) trainings, with individual runs ranging from 1 to 3.3 billion tokens each. The most successful configuration used 2.4 billion training tokens over 3 epochs. We used \href{https://github.com/arcee-ai/mergekit}{\texttt{MergeKit}}, \href{https://github.com/arcee-ai/EvolKit}{\texttt{EvolKit}}, and \href{https://github.com/arcee-ai/distillkit}{\texttt{DistillKit}} from Arcee AI to combine the models, generate the data sets, and distill the logits, respectively. For training data, we used a diverse set of high-quality datasets:

\begin{enumerate}
    \item \textbf{New Datasets} (released with \intellect):
    \begin{itemize}
        \item \texttt{arcee-ai/EvolKit-75k} (generated via \texttt{EvolKit})
        \item \texttt{arcee-ai/Llama-405B-Logits}
        \item \texttt{arcee-ai/The-Tomb}
    \end{itemize} 
    
    \item \textbf{Instruction Following}:
    \begin{itemize}
        \item \texttt{mlabonne/open-perfectblend-fixed} (generalist capabilities)
        \item \texttt{microsoft/orca-agentinstruct-1M-v1-cleaned} (Chain-of-Thought)
        \item \texttt{Post-training-Data-Flywheel/AutoIF-instruct-61k-with-funcs}
    \end{itemize}

    \item \textbf{Domain-Specific}:
    \begin{itemize}
        \item \texttt{Team-ACE/ToolACE} (function calling)
        \item Synthia coder (programming)
        \item \texttt{ServiceNow-AI/M2Lingual} (multilingual)
        \item \texttt{AI-MO/NuminaMath-TIR} (mathematics)
    \end{itemize}

    \item \textbf{Tulu-3 Persona Datasets}:
    \begin{itemize}
        \item \texttt{allenai/tulu-3-sft-personas-code}
        \item \texttt{allenai/tulu-3-sft-personas-math}
        \item \texttt{allenai/tulu-3-sft-personas-math-grade}
        \item \texttt{allenai/tulu-3-sft-personas-algebra}
    \end{itemize}
\end{enumerate}

Second, we execute 8 distinct Direct Preference Optimization (DPO) runs with various combinations of data sets to enhance specific performance metrics and align the model with human preferences. A key advantage in our post-training process was \intellect's use of the Llama-3 tokenizer, which allowed us to utilize logits from Llama-3.1-405B to heal and maintain precision during the post-training process via \texttt{DistillKit}.

Finally, we performed 16 strategic merges between candidate models using \texttt{MergeKit} to create superior combined models that leverage the strengths of different training runs. During the post-training phase, we observed that when using a ChatML template without an explicit BOS (begin-of-sequence) token, the initial loss was approximately 15. However, when switching to the Llama 3.1 chat template, the loss for these trainings started much lower at approximately 1.1, indicating better alignment with the underlying Llama 3 tokenizer.

The combination of these post-training techniques resulted in significant improvements in various benchmarks (detailed in \Cref{tab:eval_results_post_trained}), particularly in knowledge retrieval, grade school math, instruction following and reasoning. 

\subsection{Evaluation}

This section presents the main benchmark results for \intellect. We compare with similarly sized open-source models which were pre-trained in a centralized setting on similar amounts of tokens.

For the base model evaluation, we use eight benchmarks commonly featured in evaluations, such as the Hugging Face Open LLM Leaderboard. These include MMLU
5-shot~\citep{mmlu}, HellaSwag 10-shot~\citep{hellaswag}, ARC-Challenge 25-shot~\citep{arcc}, GPQA~\citep{gpqa}, GSM8K 5-shot~\citep{gsm8k}, TruthfulQA~\citep{truthfulqa}, WinoGrande 5-shot~\citep{winogrande} and Big Bench Hard (BBH)~\citep{bbh}. The results are shown in \Cref{tab:eval_results}. Additionally, for the post-trained models, we run additional benchmarks from the Hugging Face Open LLM Leaderboard 2 such as IFEval~\citep{ifeval}, and compare to other post-trained and instruction-tuned models in \cref{tab:eval_results_post_trained}. All evaluations are performed with the Language Model Evaluation Harness framework~\citep{eval-harness}. The benchmark results are promising for a 10B model trained on a trillion tokens, providing strong evidence that our DiLoCo approach effectively scales to the 10B model size.

\begin{table}[ht]
    \captionsetup{skip=6pt}
    \setlength{\tabcolsep}{4pt}
    \centering
    \begin{tabular}{lccccccc}
    \toprule
    \textbf{Model} & \textbf{Size} & \textbf{Tokens} & \textbf{MMLU} & \textbf{HellaSwag} & \textbf{ARC-C} \\
    \midrule
    \intellect  
    & 10B   & 1T    & 37.5  & 72.26 & 52.13 \\
    MPT-7B~\citep{MosaicML2023Introducing}
    & 7B    & 1T    & 26.8  & 77.41 & 46.67 \\
    Falcon-7B~\citep{falconmodel}  
    & 7B    & 1.5T  & 26.2  & 78.23 & 47.61 \\
    Pythia-12B~\citep{pythia} 
    & 12B   & 300B  & 26.5  & 68.83 & 40.61 \\
    LLM360-Amber~\citep{llm360amber}
    & 7B    & 1.3T  & 24.5  & 74.08 & 42.75 \\
    LLaMA-7B~\citep{llama1}
    & 7B    & 1T    & 35.1  & 78.19 & 50.43 \\
    LLaMA-13B~\citep{llama1}   
    & 13B   & 1T    & 46.9  & 81.05 & 56.14 \\
    LLaMA2-7B~\citep{llama2}
    & 7B    & 2T    & 45.3  & 78.64 & 54.10 \\
    LLaMA2-13B~\citep{llama2}
    & 13B   & 2T    & 54.8  & 82.58 & 59.81 \\
    
    \midrule
    & \textbf{GPQA} & \textbf{GSM8K} & \textbf{TruthfulQA} & \textbf{Winogrande} & \textbf{BBH} \\
    \midrule
    \intellect      & 26.12 & 8.1   & 35.47 & 65.82 & 32.97 \\
    MPT-7B          & 25.67 & 8.3   & 33.43 & 71.11 & 32.88 \\
    Falcon-7B       & 23.66 & 4.9   & 34.28 & 70.32 & 33.00 \\
    Pythia-12B      & 24.33 & 4.09  & 31.83 & 65.27 & 31.66 \\
    LLM360-Amber    & 27.01 & 4.32  & 40.80 & 65.35 & 31.95 \\
    LLaMA-7B        & 23.21 & 9.7   & 34.33 & 72.06 & 32.86 \\
    LLaMA-13B       & 26.34 & 17.3  & 39.48 & 76.16 & 39.74 \\
    LLaMA2-7B       & 25.89 & 13.5  & 38.75 & 74.03 & 34.46 \\
    LLaMA2-13B      & 25.67 & 24.3  & 37.38 & 77.35 & 41.68 \\

\bottomrule
\end{tabular}
\caption{Base model evaluation results across various benchmarks, measured against similarly sized open-source models pre-trained in a centralized setting on comparable amounts of total tokens.}
\label{tab:eval_results}
\end{table}

\begin{table}[ht]
\setlength{\tabcolsep}{4pt}
\captionsetup{skip=6pt}
\centering
\begin{tabular}{lccccccc}
\toprule
\textbf{Model} & \textbf{Size} & \textbf{Tokens} & \textbf{MMLU} & \textbf{HellaSwag} & \textbf{ARC-C} \\
\midrule
\intellectinstruct  & 10B   & 1T    & 49.89 & 71.42 & 54.52 \\
MPT-7B-Chat         & 7B    & 1T    & 36.29 & 75.88 & 51.02 \\
Falcon-7B-instruct  & 7B    & 1.5T  & 25.21 & 70.61 & 45.82 \\
LLM360 AmberChat    & 7B    & 1.4T  & 36.02 & 73.94 & 43.94 \\
LLaMA2-7B-chat      & 7B    & 2T    & 47.20 & 78.69 & 53.33 \\
LLaMA2-13B-chat     & 13B   & 2T    & 53.51 & 82.47 & 59.73 \\
\midrule
& \textbf{GPQA} & \textbf{GSM8K} & \textbf{TruthfulQA} & \textbf{BBH} & \textbf{IFEval} \\
\midrule
\intellectinstruct  & 28.32  & 38.58 & 42.16 & 34.85 & 40.39 \\
MPT-7B-Chat         & 26.79  & 8.26  & 35.22 & 32.30 & 14.39 \\
Falcon-7B-instruct  & 26.34  & 4.93  & 44.13 & 31.98 & 24.82 \\
LLM360 AmberChat    & 27.23  & 6.14  & 40.80 & 31.14 & 18.71 \\
LLaMA2-7B-chat      & 28.57  & 23.96 & 45.58 & 35.50 & 45.80 \\
LLaMA2-13B-chat     & 28.35  & 37.15 & 44.12 & 39.05 & 46.88 \\
\bottomrule
\end{tabular}
\caption{Post-trained model evaluation results across various benchmarks.}
\label{tab:eval_results_post_trained}
\end{table}

While \intellect\ demonstrates encouraging benchmark results and represents the first model of its size successfully trained on a decentralized network of GPUs, it still lags behind current state-of-the-art models trained on an order of magnitude more tokens. Future work will focus on scaling the model series with significantly larger compute budgets, number of contributors, introducing novel architectural advancements beyond Llama 3, and leveraging higher-quality datasets.

\section{Discussion: Decentralized Training}
\label{discussion-decentralized-ai}

The successful training of INTELLECT-1 demonstrates a key technical advancement in enabling a more open and decentralized AI ecosystem, with significant implications for the future development and governance of advanced artificial intelligence systems.

The current trajectory of AI development shows increasing consolidation of computational resources and research capabilities to a small number of corporate and state entities. As the cost of training competitive frontier models increases, this consolidation only accelerates. While centralization has the benefit of reduced coordination costs, and increased efficiency, it has the negative externality of creating significant points of control and increased risk of capability capture or misuse.

Open-source AI currently serves as the best counterweight to a closed-source AI ecosystem, and has yielded significant advancements in virtually every technical aspect of artificial intelligence development. However, open-source AI has yet to amass the scale of compute that is relevant to achieve parity with the largest closed-source AI labs. This is a requirement if the open-source ecosystem intends to train and open-source competitive frontier models. As demonstrated by this work and prior results on collaborative deep learning~\citep{lc0,diskin2021distributed,borzunov2022training}, decentralized training has the potential to enable that scale of compute by making it possible to pool resources across the world for training large-scale foundational models.

Historical precedent from distributed compute protocols and incentive networks demonstrates the potential scale of community-pooled compute resources. Bitcoin's network grew from a few personal computers to over 10 Gigawatts of compute in 12 years, and Ethereum's network reached similar scale before transitioning to proof of stake. These networks demonstrate how properly aligned economic incentives can mobilize massive computational resources across geographical and institutional boundaries. Similar dynamics could emerge in decentralized AI training: with appropriate incentive mechanisms and technical infrastructure, community-pooled AI training could potentially exceed the compute capacity of any centralized training facilities.

\section{Conclusion}
\label{conclusion}

In this report, we introduce \intellect, the first globally trained language model at the 10-billion-parameter scale.
We are open-sourcing the base model, intermediate checkpoints, pre-training data, post-training checkpoints, and post-training data.

Alongside the \intellect\ release, we are also open-sourcing \primeframework, a scalable distributed training framework designed for fault-tolerant, high-performance training on unreliable, globally distributed nodes with low network bandwidth.

We hope that this report, along with our open-source releases, will contribute to the broader ecosystem, advancing efforts to make globally distributed training a viable option for developing state-of-the-art open models.

\section*{Acknowledgements}

We would like to thank all the companies and individuals who generously provided compute resources for the training run from locations all around the world. This includes Arcee AI, @oldmankotaro, @skre\_0, @marloXBT, @rodeo\_crypro, @herb0x\_, Autonolas, @Etherean007, Hugging Face, @mev\_pete, 0xfr\_, @iroh\_pm, kiteman, realtek, primeprimeint1234, Hyperbolic, NWO, hecataeus, VirtualMachine, @notdroll, SemiAnalysis, waiting\_\_, @toptickcrypto, sto, washout\_segment\_0b and klee.

We would also like to thank Arthur Douillard for their work on DiLoCo. Tristan Rice and Junjie Wang for discussions and ideas on fault-tolerant training. Chien-Chin Huang and Iris Zhang for ideas and discussions related to asynchronous distributed checkpointing. Yifu Wang for discussions about Tensor Parallel. Andrew Gu for convincing us to switch to FSDP2 to get rid of some memory allocation issues we were facing with FSDP1.

\bibliographystyle{plainnat}
\bibliography{references}

\begin{thebibliography}{38}
\providecommand{\natexlab}[1]{#1}
\providecommand{\url}[1]{\texttt{#1}}
\expandafter\ifx\csname urlstyle\endcsname\relax
  \providecommand{\doi}[1]{doi: #1}\else
  \providecommand{\doi}{doi: \begingroup \urlstyle{rm}\Url}\fi

\bibitem[Almazrouei et~al.(2023)Almazrouei, Alobeidli, Alshamsi, Cappelli, Cojocaru, Debbah, Étienne Goffinet, Hesslow, Launay, Malartic, Mazzotta, Noune, Pannier, and Penedo]{falconmodel}
Ebtesam Almazrouei, Hamza Alobeidli, Abdulaziz Alshamsi, Alessandro Cappelli, Ruxandra Cojocaru, Mérouane Debbah, Étienne Goffinet, Daniel Hesslow, Julien Launay, Quentin Malartic, Daniele Mazzotta, Badreddine Noune, Baptiste Pannier, and Guilherme Penedo.
\newblock The falcon series of open language models, 2023.
\newblock URL \url{https://arxiv.org/abs/2311.16867}.

\bibitem[Biderman et~al.(2023)Biderman, Schoelkopf, Anthony, Bradley, O'Brien, Hallahan, Khan, Purohit, Prashanth, Raff, Skowron, Sutawika, and van~der Wal]{pythia}
Stella Biderman, Hailey Schoelkopf, Quentin Anthony, Herbie Bradley, Kyle O'Brien, Eric Hallahan, Mohammad~Aflah Khan, Shivanshu Purohit, USVSN~Sai Prashanth, Edward Raff, Aviya Skowron, Lintang Sutawika, and Oskar van~der Wal.
\newblock Pythia: A suite for analyzing large language models across training and scaling, 2023.
\newblock URL \url{https://arxiv.org/abs/2304.01373}.

\bibitem[Borzunov et~al.(2022)Borzunov, Ryabinin, Dettmers, Lhoest, Saulnier, Diskin, Jernite, and Wolf]{borzunov2022training}
Alexander Borzunov, Max Ryabinin, Tim Dettmers, Quentin Lhoest, Lucile Saulnier, Michael Diskin, Yacine Jernite, and Thomas Wolf.
\newblock Training transformers together, 2022.

\bibitem[Brown et~al.(2020)Brown, Mann, Ryder, Subbiah, Kaplan, Dhariwal, Neelakantan, Shyam, Sastry, Askell, Agarwal, Herbert{-}Voss, Krueger, Henighan, Child, Ramesh, Ziegler, Wu, Winter, Hesse, Chen, Sigler, Litwin, Gray, Chess, Clark, Berner, McCandlish, Radford, Sutskever, and Amodei]{gpt3}
Tom~B. Brown, Benjamin Mann, Nick Ryder, Melanie Subbiah, Jared Kaplan, Prafulla Dhariwal, Arvind Neelakantan, Pranav Shyam, Girish Sastry, Amanda Askell, Sandhini Agarwal, Ariel Herbert{-}Voss, Gretchen Krueger, Tom Henighan, Rewon Child, Aditya Ramesh, Daniel~M. Ziegler, Jeffrey Wu, Clemens Winter, Christopher Hesse, Mark Chen, Eric Sigler, Mateusz Litwin, Scott Gray, Benjamin Chess, Jack Clark, Christopher Berner, Sam McCandlish, Alec Radford, Ilya Sutskever, and Dario Amodei.
\newblock Language models are few-shot learners.
\newblock In Hugo Larochelle, Marc'Aurelio Ranzato, Raia Hadsell, Maria{-}Florina Balcan, and Hsuan{-}Tien Lin, editors, \emph{Advances in Neural Information Processing Systems 33: Annual Conference on Neural Information Processing Systems 2020, NeurIPS 2020, December 6-12, 2020, virtual}, 2020.
\newblock URL \url{https://proceedings.neurips.cc/paper/2020/hash/1457c0d6bfcb4967418bfb8ac142f64a-Abstract.html}.

\bibitem[Chowdhery et~al.(2022)Chowdhery, Narang, Devlin, Bosma, Mishra, Roberts, Barham, Chung, Sutton, Gehrmann, Schuh, Shi, Tsvyashchenko, Maynez, et~al.]{palm}
Aakanksha Chowdhery, Sharan Narang, Jacob Devlin, Maarten Bosma, Gaurav Mishra, Adam Roberts, Paul Barham, Hyung~Won Chung, Charles Sutton, Sebastian Gehrmann, Parker Schuh, Kensen Shi, Sasha Tsvyashchenko, Joshua Maynez, et~al.
\newblock Palm: Scaling language modeling with pathways, 2022.
\newblock URL \url{https://arxiv.org/abs/2204.02311}.

\bibitem[Clark et~al.(2018)Clark, Cowhey, Etzioni, Khot, Sabharwal, Schoenick, and Tafjord]{arcc}
Peter Clark, Isaac Cowhey, Oren Etzioni, Tushar Khot, Ashish Sabharwal, Carissa Schoenick, and Oyvind Tafjord.
\newblock Think you have solved question answering? try arc, the ai2 reasoning challenge, 2018.
\newblock URL \url{https://arxiv.org/abs/1803.05457}.

\bibitem[Cobbe et~al.(2021)Cobbe, Kosaraju, Bavarian, Chen, Jun, Kaiser, Plappert, Tworek, Hilton, Nakano, Hesse, and Schulman]{gsm8k}
Karl Cobbe, Vineet Kosaraju, Mohammad Bavarian, Mark Chen, Heewoo Jun, Lukasz Kaiser, Matthias Plappert, Jerry Tworek, Jacob Hilton, Reiichiro Nakano, Christopher Hesse, and John Schulman.
\newblock Training verifiers to solve math word problems, 2021.
\newblock URL \url{https://arxiv.org/abs/2110.14168}.

\bibitem[Dettmers et~al.(2022)Dettmers, Lewis, Belkada, and Zettlemoyer]{llmint8}
Tim Dettmers, Mike Lewis, Younes Belkada, and Luke Zettlemoyer.
\newblock Llm.int8(): 8-bit matrix multiplication for transformers at scale, 2022.
\newblock URL \url{https://arxiv.org/abs/2208.07339}.

\bibitem[Diskin et~al.(2021)Diskin, Bukhtiyarov, Ryabinin, Saulnier, Lhoest, Sinitsin, Popov, Pyrkin, Kashirin, Borzunov, del Moral, Mazur, Kobelev, Jernite, Wolf, and Pekhimenko]{diskin2021distributed}
Michael Diskin, Alexey Bukhtiyarov, Max Ryabinin, Lucile Saulnier, Quentin Lhoest, Anton Sinitsin, Dmitry Popov, Dmitriy Pyrkin, Maxim Kashirin, Alexander Borzunov, Albert~Villanova del Moral, Denis Mazur, Ilia Kobelev, Yacine Jernite, Thomas Wolf, and Gennady Pekhimenko.
\newblock Distributed deep learning in open collaborations.
\newblock In \emph{Advances in Neural Information Processing Systems}, 2021.
\newblock URL \url{https://openreview.net/forum?id=FYHktcK-7v}.

\bibitem[Douillard et~al.(2024)Douillard, Feng, Rusu, Chhaparia, Donchev, Kuncoro, Ranzato, Szlam, and Shen]{diloco}
Arthur Douillard, Qixuan Feng, Andrei~A. Rusu, Rachita Chhaparia, Yani Donchev, Adhiguna Kuncoro, Marc'Aurelio Ranzato, Arthur Szlam, and Jiajun Shen.
\newblock Diloco: Distributed low-communication training of language models, 2024.
\newblock URL \url{https://arxiv.org/abs/2311.08105}.

\bibitem[Gao et~al.(2024)Gao, Tow, Abbasi, Biderman, Black, DiPofi, Foster, Golding, Hsu, Le~Noac'h, Li, McDonell, Muennighoff, Ociepa, Phang, Reynolds, Schoelkopf, Skowron, Sutawika, Tang, Thite, Wang, Wang, and Zou]{eval-harness}
Leo Gao, Jonathan Tow, Baber Abbasi, Stella Biderman, Sid Black, Anthony DiPofi, Charles Foster, Laurence Golding, Jeffrey Hsu, Alain Le~Noac'h, Haonan Li, Kyle McDonell, Niklas Muennighoff, Chris Ociepa, Jason Phang, Laria Reynolds, Hailey Schoelkopf, Aviya Skowron, Lintang Sutawika, Eric Tang, Anish Thite, Ben Wang, Kevin Wang, and Andy Zou.
\newblock A framework for few-shot language model evaluation, 07 2024.
\newblock URL \url{https://zenodo.org/records/12608602}.

\bibitem[Grattafiori et~al.(2024)Grattafiori, Dubey, Jauhri, Pandey, Kadian, Al-Dahle, Letman, Mathur, Schelten, Vaughan, Yang, Fan, et~al.]{llama3}
Aaron Grattafiori, Abhimanyu Dubey, Abhinav Jauhri, Abhinav Pandey, Abhishek Kadian, Ahmad Al-Dahle, Aiesha Letman, Akhil Mathur, Alan Schelten, Alex Vaughan, Amy Yang, Angela Fan, et~al.
\newblock The llama 3 herd of models, 2024.
\newblock URL \url{https://arxiv.org/abs/2407.21783}.

\bibitem[Hendrycks et~al.(2021)Hendrycks, Burns, Basart, Zou, Mazeika, Song, and Steinhardt]{mmlu}
Dan Hendrycks, Collin Burns, Steven Basart, Andy Zou, Mantas Mazeika, Dawn Song, and Jacob Steinhardt.
\newblock Measuring massive multitask language understanding, 2021.
\newblock URL \url{https://arxiv.org/abs/2009.03300}.

\bibitem[Hägele et~al.(2024)Hägele, Bakouch, Kosson, Allal, Werra, and Jaggi]{hägele2024scalinglawscomputeoptimaltraining}
Alexander Hägele, Elie Bakouch, Atli Kosson, Loubna~Ben Allal, Leandro~Von Werra, and Martin Jaggi.
\newblock Scaling laws and compute-optimal training beyond fixed training durations, 2024.
\newblock URL \url{https://arxiv.org/abs/2405.18392}.

\bibitem[Jaghouar et~al.(2024)Jaghouar, Ong, and Hagemann]{open_diloco}
Sami Jaghouar, Jack~Min Ong, and Johannes Hagemann.
\newblock Opendiloco: An open-source framework for globally distributed low-communication training, 2024.
\newblock URL \url{https://arxiv.org/abs/2407.07852}.

\bibitem[Kocetkov et~al.(2022)Kocetkov, Li, Allal, Li, Mou, Ferrandis, Jernite, Mitchell, Hughes, Wolf, Bahdanau, von Werra, and de~Vries]{thestack}
Denis Kocetkov, Raymond Li, Loubna~Ben Allal, Jia Li, Chenghao Mou, Carlos~Muñoz Ferrandis, Yacine Jernite, Margaret Mitchell, Sean Hughes, Thomas Wolf, Dzmitry Bahdanau, Leandro von Werra, and Harm de~Vries.
\newblock The stack: 3 tb of permissively licensed source code, 2022.
\newblock URL \url{https://arxiv.org/abs/2211.15533}.

\bibitem[Kumar et~al.(2024)Kumar, Ankner, Spector, Bordelon, Muennighoff, Paul, Pehlevan, Ré, and Raghunathan]{scaling_law_quant}
Tanishq Kumar, Zachary Ankner, Benjamin~F. Spector, Blake Bordelon, Niklas Muennighoff, Mansheej Paul, Cengiz Pehlevan, Christopher Ré, and Aditi Raghunathan.
\newblock Scaling laws for precision, 2024.
\newblock URL \url{https://arxiv.org/abs/2411.04330}.

\bibitem[Li et~al.(2024)Li, Fang, Smyrnis, Ivgi, Jordan, Gadre, Bansal, Guha, Keh, Arora, Garg, Xin, Muennighoff, Heckel, Mercat, Chen, Gururangan, Wortsman, Albalak, et~al.]{dclm}
Jeffrey Li, Alex Fang, Georgios Smyrnis, Maor Ivgi, Matt Jordan, Samir Gadre, Hritik Bansal, Etash Guha, Sedrick Keh, Kushal Arora, Saurabh Garg, Rui Xin, Niklas Muennighoff, Reinhard Heckel, Jean Mercat, Mayee Chen, Suchin Gururangan, Mitchell Wortsman, Alon Albalak, et~al.
\newblock Datacomp-lm: In search of the next generation of training sets for language models, 2024.
\newblock URL \url{https://arxiv.org/abs/2406.11794}.

\bibitem[Lin et~al.(2022)Lin, Hilton, and Evans]{truthfulqa}
Stephanie Lin, Jacob Hilton, and Owain Evans.
\newblock Truthfulqa: Measuring how models mimic human falsehoods, 2022.
\newblock URL \url{https://arxiv.org/abs/2109.07958}.

\bibitem[Liu et~al.(2023)Liu, Qiao, Neiswanger, Wang, Tan, Tao, Li, Wang, Sun, Pangarkar, Fan, Gu, Miller, Zhuang, He, Li, Koto, Tang, Ranjan, Shen, Ren, Iriondo, Mu, Hu, Schulze, Nakov, Baldwin, and Xing]{llm360amber}
Zhengzhong Liu, Aurick Qiao, Willie Neiswanger, Hongyi Wang, Bowen Tan, Tianhua Tao, Junbo Li, Yuqi Wang, Suqi Sun, Omkar Pangarkar, Richard Fan, Yi~Gu, Victor Miller, Yonghao Zhuang, Guowei He, Haonan Li, Fajri Koto, Liping Tang, Nikhil Ranjan, Zhiqiang Shen, Xuguang Ren, Roberto Iriondo, Cun Mu, Zhiting Hu, Mark Schulze, Preslav Nakov, Tim Baldwin, and Eric~P. Xing.
\newblock Llm360: Towards fully transparent open-source llms, 2023.
\newblock URL \url{https://arxiv.org/abs/2312.06550}.

\bibitem[OpenAI(2023)]{gpt4}
OpenAI.
\newblock Gpt-4 technical report, 2023.

\bibitem[Pascutto and Linscott(2019)]{lc0}
Gian-Carlo Pascutto and Gary Linscott.
\newblock Leela chess zero, 2019.
\newblock URL \url{http://lczero.org/}.

\bibitem[Paster et~al.(2023)Paster, Santos, Azerbayev, and Ba]{openwebmath}
Keiran Paster, Marco~Dos Santos, Zhangir Azerbayev, and Jimmy Ba.
\newblock Openwebmath: An open dataset of high-quality mathematical web text, 2023.
\newblock URL \url{https://arxiv.org/abs/2310.06786}.

\bibitem[Penedo et~al.(2024)Penedo, Kydlíček, allal, Lozhkov, Mitchell, Raffel, Werra, and Wolf]{fineweb}
Guilherme Penedo, Hynek Kydlíček, Loubna~Ben allal, Anton Lozhkov, Margaret Mitchell, Colin Raffel, Leandro~Von Werra, and Thomas Wolf.
\newblock The fineweb datasets: Decanting the web for the finest text data at scale, 2024.
\newblock URL \url{https://arxiv.org/abs/2406.17557}.

\bibitem[Rein et~al.(2023)Rein, Hou, Stickland, Petty, Pang, Dirani, Michael, and Bowman]{gpqa}
David Rein, Betty~Li Hou, Asa~Cooper Stickland, Jackson Petty, Richard~Yuanzhe Pang, Julien Dirani, Julian Michael, and Samuel~R. Bowman.
\newblock {GPQA: A Graduate-Level Google-Proof Q\&A Benchmark}, 2023.
\newblock URL \url{https://arxiv.org/abs/2311.12022}.

\bibitem[Ryabinin and Gusev(2020)]{ryabinin2020crowdsourced}
Max Ryabinin and Anton Gusev.
\newblock Towards crowdsourced training of large neural networks using decentralized mixture-of-experts.
\newblock In \emph{Advances in Neural Information Processing Systems}, volume~33, 2020.
\newblock URL \url{https://proceedings.neurips.cc/paper/2020/file/25ddc0f8c9d3e22e03d3076f98d83cb2-Paper.pdf}.

\bibitem[Ryabinin et~al.(2020)Ryabinin, Borzunov, Diskin, Gusev, Mazur, Plokhotnyuk, Bukhtiyarov, Samygin, Sinitsin, and Chumachenko]{hivemind}
Max Ryabinin, Alexander Borzunov, Michael Diskin, Anton Gusev, Denis Mazur, Vsevolod Plokhotnyuk, Alexey Bukhtiyarov, Pavel Samygin, Anton Sinitsin, and Artem Chumachenko.
\newblock {H}ivemind: {D}ecentralized {D}eep {L}earning in {P}y{T}orch, April 2020.
\newblock URL \url{https://github.com/learning-at-home/hivemind}.

\bibitem[Sakaguchi et~al.(2019)Sakaguchi, Bras, Bhagavatula, and Choi]{winogrande}
Keisuke Sakaguchi, Ronan~Le Bras, Chandra Bhagavatula, and Yejin Choi.
\newblock Winogrande: An adversarial winograd schema challenge at scale, 2019.
\newblock URL \url{https://arxiv.org/abs/1907.10641}.

\bibitem[Stich(2019)]{stich2018local}
Sebastian~U. Stich.
\newblock Local {SGD} converges fast and communicates little.
\newblock In \emph{International Conference on Learning Representations}, 2019.
\newblock URL \url{https://openreview.net/forum?id=S1g2JnRcFX}.

\bibitem[Suzgun et~al.(2022)Suzgun, Scales, Schärli, Gehrmann, Tay, Chung, Chowdhery, Le, Chi, Zhou, and Wei]{bbh}
Mirac Suzgun, Nathan Scales, Nathanael Schärli, Sebastian Gehrmann, Yi~Tay, Hyung~Won Chung, Aakanksha Chowdhery, Quoc~V. Le, Ed~H. Chi, Denny Zhou, and Jason Wei.
\newblock Challenging big-bench tasks and whether chain-of-thought can solve them, 2022.
\newblock URL \url{https://arxiv.org/abs/2210.09261}.

\bibitem[Team(2023)]{MosaicML2023Introducing}
MosaicML~NLP Team.
\newblock Introducing mpt-7b: A new standard for open-source, commercially usable llms, 2023.
\newblock URL \url{www.mosaicml.com/blog/mpt-7b}.
\newblock Accessed: 2023-05-05.

\bibitem[Thakur et~al.(2005)Thakur, Rabenseifner, and Gropp]{allreduce}
Rajeev Thakur, Rolf Rabenseifner, and William Gropp.
\newblock Optimization of collective communication operations in mpich.
\newblock \emph{The International Journal of High Performance Computing Applications}, 19:\penalty0 49 -- 66, 2005.
\newblock URL \url{https://api.semanticscholar.org/CorpusID:90404}.

\bibitem[Touvron et~al.(2023{\natexlab{a}})Touvron, Lavril, Izacard, Martinet, Lachaux, Lacroix, Rozière, Goyal, Hambro, Azhar, Rodriguez, Joulin, Grave, and Lample]{llama1}
Hugo Touvron, Thibaut Lavril, Gautier Izacard, Xavier Martinet, Marie-Anne Lachaux, Timothée Lacroix, Baptiste Rozière, Naman Goyal, Eric Hambro, Faisal Azhar, Aurelien Rodriguez, Armand Joulin, Edouard Grave, and Guillaume Lample.
\newblock Llama: Open and efficient foundation language models, 2023{\natexlab{a}}.
\newblock URL \url{https://arxiv.org/abs/2302.13971}.

\bibitem[Touvron et~al.(2023{\natexlab{b}})Touvron, Martin, Stone, Albert, Almahairi, Babaei, Bashlykov, Batra, Bhargava, Bhosale, Bikel, Blecher, Ferrer, et~al.]{llama2}
Hugo Touvron, Louis Martin, Kevin Stone, Peter Albert, Amjad Almahairi, Yasmine Babaei, Nikolay Bashlykov, Soumya Batra, Prajjwal Bhargava, Shruti Bhosale, Dan Bikel, Lukas Blecher, Cristian~Canton Ferrer, et~al.
\newblock Llama 2: Open foundation and fine-tuned chat models, 2023{\natexlab{b}}.
\newblock URL \url{https://arxiv.org/abs/2307.09288}.

\bibitem[Yang et~al.(2023)Yang, Xiao, Wang, Zhang, Bian, Yin, Lv, Pan, Wang, Yan, Yang, Deng, Wang, Liu, Ai, Dong, Zhao, Xu, Sun, Zhang, Liu, Ji, Xie, Dai, Fang, Su, Song, Liu, Ru, Ma, Wang, Liu, Lin, Nie, Guo, Sun, Zhang, Li, Li, Cheng, Chen, Zeng, Wang, Chen, Men, Yu, Pan, Shen, Wang, Li, Jiang, Gao, Zhang, Zhou, and Wu]{baichuan2}
Aiyuan Yang, Bin Xiao, Bingning Wang, Borong Zhang, Ce~Bian, Chao Yin, Chenxu Lv, Da~Pan, Dian Wang, Dong Yan, Fan Yang, Fei Deng, Feng Wang, Feng Liu, Guangwei Ai, Guosheng Dong, Haizhou Zhao, Hang Xu, Haoze Sun, Hongda Zhang, Hui Liu, Jiaming Ji, Jian Xie, JunTao Dai, Kun Fang, Lei Su, Liang Song, Lifeng Liu, Liyun Ru, Luyao Ma, Mang Wang, Mickel Liu, MingAn Lin, Nuolan Nie, Peidong Guo, Ruiyang Sun, Tao Zhang, Tianpeng Li, Tianyu Li, Wei Cheng, Weipeng Chen, Xiangrong Zeng, Xiaochuan Wang, Xiaoxi Chen, Xin Men, Xin Yu, Xuehai Pan, Yanjun Shen, Yiding Wang, Yiyu Li, Youxin Jiang, Yuchen Gao, Yupeng Zhang, Zenan Zhou, and Zhiying Wu.
\newblock Baichuan 2: Open large-scale language models, 2023.
\newblock URL \url{https://arxiv.org/abs/2309.10305}.

\bibitem[Zellers et~al.(2019)Zellers, Holtzman, Bisk, Farhadi, and Choi]{hellaswag}
Rowan Zellers, Ari Holtzman, Yonatan Bisk, Ali Farhadi, and Yejin Choi.
\newblock Hellaswag: Can a machine really finish your sentence?, 2019.
\newblock URL \url{https://arxiv.org/abs/1905.07830}.

\bibitem[Zhao et~al.(2023)Zhao, Gu, Varma, Luo, Huang, Xu, Wright, Shojanazeri, Ott, Shleifer, Desmaison, Balioglu, Damania, Nguyen, Chauhan, Hao, Mathews, and Li]{fsdp}
Yanli Zhao, Andrew Gu, Rohan Varma, Liang Luo, Chien-Chin Huang, Min Xu, Less Wright, Hamid Shojanazeri, Myle Ott, Sam Shleifer, Alban Desmaison, Can Balioglu, Pritam Damania, Bernard Nguyen, Geeta Chauhan, Yuchen Hao, Ajit Mathews, and Shen Li.
\newblock Pytorch fsdp: Experiences on scaling fully sharded data parallel, 2023.
\newblock URL \url{https://arxiv.org/abs/2304.11277}.

\bibitem[Zhou et~al.(2023)Zhou, Lu, Mishra, Brahma, Basu, Luan, Zhou, and Hou]{ifeval}
Jeffrey Zhou, Tianjian Lu, Swaroop Mishra, Siddhartha Brahma, Sujoy Basu, Yi~Luan, Denny Zhou, and Le~Hou.
\newblock Instruction-following evaluation for large language models, 2023.
\newblock URL \url{https://arxiv.org/abs/2311.07911}.

\end{thebibliography}


\appendix

\newpage
\section{Model Configuration}

\label{appendix:model-configuration}

\begin{table}[ht]
\caption{Model Configuration for \intellect. The model is based on the Llama3 architecture~\citep{llama3}, with the only difference from Llama3-8B being the number of layers.}
\label{tab:model_config}
\vskip 0.15in
\begin{center}
\begin{small}
\begin{sc}
\begin{tabular}{lc}
\toprule
Parameter & 10B \\
\midrule
Number of layers & $42$ \\
Hidden dim & $4{\small,}096$ \\
Number of heads & $32$ \\
K/V size & $8$ \\
Vocab size & $128{\small,}256$ \\
Inner learning rate & $7.5e-5$ \\
Warmup steps & $1{\small,}000$ \\
Weight decay & $0.1$ \\
Batch Size & $128$ \\
Sequence length & $8{\small,}192$ \\
Outer Nesterov LR & $0.7$ \\
Outer Nesterov momentum & $0.9$ \\
Max Z Loss weight  & $2e-4$ \\
Adam Beta 1 & 0.9 \\
Adam Beta 2 & 0.95 \\
Weight Decay & 0.1 \\
\bottomrule
\end{tabular}
\end{sc}
\end{small}
\end{center}
\vskip -0.1in
\end{table}

\newpage

\end{document}